\pdfoutput=1

\documentclass{article}
\usepackage{ijcai20}
\usepackage{color, colortbl}

\usepackage{times}
\usepackage{soul}
\usepackage{url}
\usepackage[hidelinks]{hyperref}
\usepackage[utf8]{inputenc}
\usepackage[small]{caption}
\usepackage{graphicx}
\usepackage{amsmath}
\usepackage{amsthm}
\usepackage{booktabs}
\usepackage{algorithm}
\usepackage{algorithmic}
\usepackage{natbib}

\title{OPRA: An Open-Source Online Preference Reporting and Aggregation System\\
\Large Video at: youtube.com/watch?v=O56L233rr84}

\author{
Yiwei Chen$^1$\and
Jingwen Qian$^1$\and
Junming Wang$^2$\and
Lirong Xia$^1$\footnote{Contact Author}\And
Gavriel Zahavi$^1$
\\
\affiliations
$^1$Rensselaer Polytechnic Institute \ \ \
$^2$Stanford University\\
\emails
reedyka3@gmail.com,
qianj2@rpi.edu,
jumingw@cs.stanford.edu,
xial@cs.rpi.edu,
gavriel.zahavi@gmail.com
}

\begin{document}

\maketitle

\begin{abstract}
  We introduce the Online Preference Reporting and Aggregation (OPRA) system, an open-source online system that aims at providing support for group decision-making. We illustrate OPRA's distinctive features: UI for reporting rankings with ties, comprehensive analytics of preferences, and group decision-making in combinatorial domains. We also discuss our work in an automatic mentor matching system. We hope that the open-source nature of OPRA will foster  development of computerized group decision support systems.
\end{abstract}

\section{Introduction}
The field of {\em social choice}, sometimes referred to as {\em group decision-making} or {\em collective decision-making}, aims at designing mechanisms to help people make a joint decision, despite that they may have conflicting preferences. Typical examples of social choice are (1) {\bf voting}, ranging from high-stakes (e.g. presidential elections) to low-stakes (e.g. deciding a restaurant for dinner) voting scenarios; and (2) {\bf resource allocation}, ranging from high-stakes (e.g. allocating resources among countries) to low-stakes (allocating course projects to students).

In this paper, we introduce a system for online group decision-making: the Online Preference Reporting and Aggregation (OPRA) system,\footnote{site: opra.io; source code: github.com/PrefPy/opra} which has the following distinctive features:

\noindent  {\bf User interfaces for reporting rankings with ties.} OPRA provides 5 intuitive interfaces for users to rank the alternatives with ties. In particular, the {\em one-column} and {\em two-column} UIs allow users to report rankings with ties via drag-and-drop operations.

\noindent    {\bf Comprehensive analytics of preferences.} In addition to standard information and statistics about voting outcomes, OPRA also computes and displays other consensus metrics, such as clusters of preferences based on learning mixture of Plackett-Luce models~\citep{Zhao16:Learning}, and margin of victory~\citep{Xia12:Computing}.

\noindent   {\bf Group decision-making in combinatorial domains.} OPRA supports reporting of Conditional Preference Networks (CP-nets)~\citep{Boutilier04:CP} for multi-issue voting~\citep{Lang16:Voting} and multi-type resource allocation~\citep{Mackin2016:Allocating}.

\noindent    {\bf Matching} OPRA includes a system for automatic assignment of student mentors to courses based on their preferences and qualifications.

\noindent    {\bf Open-source.} OPRA is open-source, allowing easy modification and customization for deployment, and easy adaptation of its components to other systems.



\section{Voting UIs}
\begin{figure}[h!]
	\centering
	\includegraphics[width=\linewidth]{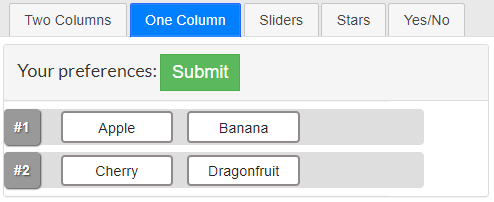}
	\caption{\label{fig:ui} \small OPRA's one column voting UI.}
\end{figure}

OPRA includes the following five UIs for users to report weak orders over alternatives. (1)    \textbf{One-Column (Figure \ref{fig:ui}).} OPRA's one-column UI  is an extension of the widely-used JQuery sortable class.
Users perform drag-and-drop operations to achieve the rank order they desire~\citep{li2019minimizing}. This UI supports ties among alternatives.(2)    \textbf{Two-Column.} The two-column UI is a variation of the one-column UI. The left column is the same as one-column, and the right column allows clicking operation to rank. the alternatives that remain in the right column are submitted as unranked. (3) \textbf{Sliders.} each alternative is ranked from 0 to 100 using a slider. (4)   \textbf{Star Rating.} each alternative is ranked from 0 to 10 using a star rating UI. (5)   \textbf{Yes/No.} each alternative is associated with a checkbox allowing submission of approval vote.

\section{Comprehensive Analytics of Preferences}

OPRA computes and shows the results of some voting rules, preference learning, and margin of victory.

\noindent{\bf Commonly-studied voting rules. }
OPRA implements commonly-studied voting rules, including { Plurality}, {Borda}, {$k$-Approval}, {Single Transferable Vote (STV)}, and others. For a poll, each voting rule is applied. The result is displayed in a table, and the winner can be seen under different voting rules (Figure \ref{fig:result} below shows some rows from an example table). While most voting rules are implemented using standard algorithms based on their definitions, some voting rules' implementation uses state-of-the-art algorithms proposed in recent literature. For example, implementation of STV and Ranked pairs computes all parallel universe tie-breaking (PUT) winners ~\citep{Conitzer09:Preference} using a DFS-based algorithm in~\citep{Wang2019:Practical}.

\begin{figure}[h!]
	\centering
	\includegraphics[width=\linewidth]{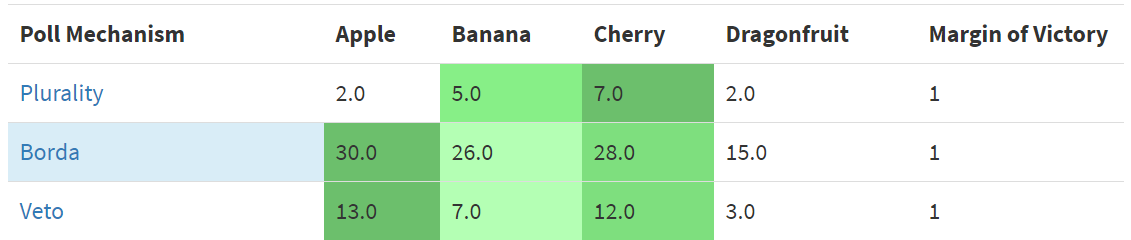}
	\caption{\label{fig:result} \small An example of poll result table on OPRA. Here, Cherry wins under Plurality, but Apple wins under Borda and Veto.}
\end{figure}

\noindent{\bf Margin of victory.} OPRA computes and shows the {\em margin of victory}~\citep{Magrino11:Computing,Cary11:Estimating,Xia12:Computing}. Margin of victory measures the robustness of the voting outcome, and also plays an important role in post-election audits. The higher the margin of victory is, the more robust the outcome is.

\noindent{\bf Preference learning.} OPRA implements the EGMM method~\citep{Zhao16:Learning} of learning mixtures of Plackett-Luce model. Learned mixtures offer a high-level overview of groups and strength of preferences based on users' ordinal preferences and can be used to make useful decisions such as group activity selection~\citep{Darmann2012:Group}.

\section{Group Decision-Making in Combinatorial Domains}
In a combinatorial domain, there are exponentially many alternatives, each of which is characterized by its values on $p$ variables. For example, in combinatorial voting~\citep{Lang16:Voting}, voters vote to decide approval/non-approval of $p$ issues. In multi-type resource allocation~\citep{Mackin2016:Allocating}, there are $p$ types of items to be allocated to agents according to their preferences.

 The challenges in group decision-making in combinatorial domains are: {\bf First}, there are often exponentially many alternatives/bundles of items, which makes it hard for users to report their preferences. {\bf Second}, it is unclear what mechanisms should be applied when users use a compact language, e.g.~CP-nets as we will recall shortly, to represent their preferences.
See~\citep{Lang16:Voting} for more discussions.


To address the first challenge, OPRA allows users submit CP-nets~\citep{Boutilier04:CP}. To address the second challenge, OPRA supports {\em sequential voting}~\citep{Lang07:Vote} on multiple issues and {\em sequential resource allocation} of multiple types of items, via the ``Multi-Poll" section.

To execute sequential voting, OPRA allows users to express preferences either using a CP-net, indicating their conditional preferences, or by waiting until previous sub-polls finish, then submit their preferences over the current issue.

For sequential resource allocation, OPRA allows users to use {\em serial dictatorships} to allocate indivisible items, when the number of users is the same as the number of items.

\section{Matching}
OPRA offers an online Mentor application tool to automatically assign student mentors to courses looking for mentors. The matching is based on the kinds of mentors a course is looking for, and the courses desired for mentoring by a student. For each course, instructors and administrators are able to modify the weights of its features (e.g. how important are GPA, mentor experience, etc.), while from the student's side, they can submit their application form and their rankings of courses. The personal features and course preferences will be taken as their input of the matching algorithm. Administrators can also modify weights, add, remove, or fix students through the ranking UI, and perform real-time re-matching.

\noindent  {\bf Matching algorithm.} For a given course, we calculate a student's score for that course as the dot product between the course's weight vector and the student's feature vector. this is done for each student, the resulting scores giving a preference ranking for this course over all the students. we then use these courses' rankings over students and students' (partial) rankings over courses to perform a course-proposing Gale-Shapley stable matching algorithm \citep{Gale62:College} (with courses also having a max capacity). This process can be re-run by administrators at any time.

\begin{figure}[h!]
	\centering
	\includegraphics[width=\linewidth]{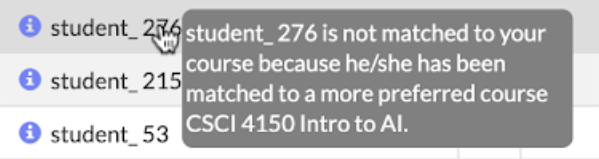}
	\caption{\label{fig:explainableai} \small An example of Explainable AI on OPRA.}
\end{figure}

\noindent   {\bf Explainable AI.} The system also supports some Explainable AI functionality \citep{ribera2019can}. From their page, administrators can view reasons why a given student was a assigned to a class or not. For example, a student wasn't assigned to the class because they were assigned to a higher-ranked class instead, as seen in Figure \ref{fig:explainableai}.

\section{Future Work}
\begin{itemize}
    \item {\bf Usability:} Improve human computer interaction, such as offering initial rankings to users without introducing bias.
    \item {\bf Explainability:} Add more comprehensive analytics and statistics for the purpose of better explanations.
    \item {\bf Algorithms:} Improve algorithms for polls to aid in group activity selection and group decision making, as well as improve matching algorithms to handle min-max bounds to course capacity \citep{nasre2017popular}.
\end{itemize}

\newpage
\section{Acknowledgments}

This work is supported by NSF \#1453542, NSF \#1716333, and ONR \#N00014-17-1-2621.

\bibliographystyle{named}
\bibliography{references}

\end{document}